\documentclass[aps,10pt,showpacs,showkeys,superscriptaddress]{revtex4}

\usepackage{amsmath}
\usepackage{amssymb}
\usepackage{amsfonts}
\usepackage{revsymb}
\usepackage{graphicx}

\begin{document}

\title{Microfield Fluctuations and Spectral Line Shapes in Strongly Coupled Two--Component Plasmas}
\author{H. B. Nersisyan}
\email{hrachya@irphe.am}
\affiliation{Division of Theoretical Physics, Institute of Radiophysics and
Electronics, Alikhanian Brothers Street 1, 378410 Ashtarak, Armenia}

\author{C. Toepffer}
\author{G. Zwicknagel}
\email{guenter.zwicknagel@physik.uni-erlangen.de}
\affiliation{Institut f\"{u}r Theoretische Physik, Department Physik,
Universit\"{a}t Erlangen-N\"{u}rnberg, Staudtstr. 7, D-91058 Erlangen, Germany}

\begin{abstract}
The spectral line shapes for hydrogen--like heavy ion emitters embedded in
strongly correlated two--component electron--ion plasmas are investigated with
numerical simulations. For that purpose the microfield fluctuations are
calculated by molecular dynamics simulations where short range quantum effects
are taken into account by using a regularized Coulomb potential for the
electron--ion interaction. The microfield fluctuations are used as input in
a numerical solution of the time--dependent Schr\"{o}dinger equation for the
radiating electron. In distinction to the standard impact and quasistatic
approximations the method presented here allows to account for the correlations
between plasma ions and electrons. The shapes of the Ly$_{\alpha}$--line in Al
are investigated in the intermediate regime. The calculations are in good
agreement with experiments on the Ly$_{\alpha}$--line in laser generated
plasmas.
\end{abstract}


\maketitle

\section{Introduction}
\label{sec:1}

Measurements of emission and absorption spectra of atoms and ions are one
of the most important tools in plasma diagnostics \cite{gri97,sal98,dat}.
They allow for investigations of the properties of various laboratory and
space plasmas. In particular, spectral line shapes may be analyzed to yield
a wealth of information on the plasma parameters provided, however, that
the data are compared to accurate computations of the spectral line broadening.

Traditionally, in the theory of spectral line broadening in plasmas ion
effects in most cases were calculated within the quasistatic approximation,
while the electron perturbation was believed to satisfy the impact approximation.
The description of this method (Standard Theory (ST)) is given, e.g., in
Refs.~\cite{gri97,sal98,dat}. The separation
of perturbations into ion and electron parts, in general, cannot be made
without a loss of accuracy, although it is argued \cite{ale96} that in many
cases it is justified. A more serious problem, however, is that each of the
ion and electron parts often needs to be considered beyond the limits of the
quasistatic and impact approximations. In particular, the ion motion in plasma
leads to the so--called ion dynamics effects. It was first shown theoretically
\cite{duf69,fri71} and soon found in experiments \cite{kel73,gru77,gru78}, that
the ion dynamics can be responsible for significant corrections to the spectral
line widths. In order to advance the calculations beyond the ST, several numerical
methods have been developed. Among the first is the method developed in
Ref.~\cite{sta83}, where a computer code was used to
simulate the ion motion along straight paths, while the electron contribution
was calculated using the impact approximation. The method was further
improved in Ref.~\cite{sta86} by using molecular--dynamics (MD) simulations
for the ions, thus accounting for interactions between the radiators and the
ion perturbers. In Refs.~\cite{gig87,heg88,car89,gig96} the motion of both ions
and electrons was numerically simulated. The particle motion was simulated using
straight path trajectories, which is applicable when the correlations between
the perturbers and radiator are neglected. Later, the area of applicability
was extended by using hyperbolic \cite{ale97} and, recently, exact paths for
the perturbers (see, e.g.,
Refs.~\cite{mar03,mar04,sta06} and references therein). In Refs.~\cite{mar03,mar04}
the model was based on one--component plasmas (OCPs) treating the full
spectrum as a superposition of the electronic and ionic contributions and thus
neglects the influence of the attractive interactions between electrons and ions.
This is well justified for weakly coupled plasma where the ionic and electronic
fields can be handled separately for spectra modeling. But there is an increasing
number of experiments of interest which are far beyond such parameter regimes
(see, e.g., Refs.~\cite{sae99,eid00,and02,eid03}). In such cases a simple superposition
of the electronic and ionic fields becomes insufficient due to strong nonlinear
effects and the total field in a two--component plasmas (TCP) should serve as
the starting point for spectra modeling. The related microfield distribution
(MFD) and the line shapes including the full attractive electron--ion interaction
has thus been attracting more and more attention and has already been studied,
e.g., in Refs.~\cite{sad09,ner05,ner06,ner08,cal07,fer07,sta07,tal02}.

The present paper is the continuation of Ref.~\cite{mar04} but we treat here
the ions and electrons on an equal footing by concentrating on the TCP. For that
purpose we perform MD simulations which span the entire range between the impact
and the quasistatic approximations. We solve the time--dependent Schr\"{o}dinger
equation for the radiator in the fluctuating microfield generated by the plasma
particles. The MD simulations in conjunction with the Schr\"{o}dinger equation to
study the line shapes in a TCP have been previously considered in Ref.~\cite{sta06}.
Our model is thus similar to that considered in \cite{sta06} but, in addition,
it allows to treat the interaction of the emitted photons with radiating electron.

\section{MD simulations}
\label{sec:2}

A two--component classical plasma of electrons and ions (with the charge $Ze$) is
in equilibrium with the temperature $T$ completely described by the coupling
parameters $\Gamma _{\alpha \beta }$ with $\alpha $,$\beta =e$,$i$. Introducing
the mean electron--electron ($a_{e}$), ion--ion ($a_{i}$) and electron--ion ($a$)
distances through the relations, $a_{\alpha}^{-3}=4\pi n_{\alpha}/3$, $a^{-3}=4\pi n/3$
(where $n=n_{e}+n_{i}$ is the total plasma density with $n_{e}=Zn_{i}$) these
parameters are defined as~\cite{ner05,ner06,ner08}
\begin{equation}
\Gamma _{\alpha\alpha}=\frac{q^{2}_{\alpha} e_{S}^{2}}{k_{B}T a_{\alpha}} \ ,
\quad  \Gamma _{ei}=\frac{Ze_{S}^{2}}{k_{B}T a} \ .
\label{eq:11}
\end{equation}%
Here $q_{e}=-1$, $q_{i}=Z$, $e_{S}^{2}=e^{2}/4\pi \varepsilon _{0}$ and $\varepsilon _{0}$
is the permittivity of the vacuum.

It is well known \cite{kel63,deu81} that, to avoid
the collapse of the classical system of electrons and ions, the Coulomb electron--ion
interaction potential must be replaced by the pseudopotential with a regularized
short--range behavior. In this paper we consider the electron--ion pair interaction
potential $-e_{S}^{2} q_{\beta}u_{ei}(r)$, where $\beta =i,R$, $q_{R}e$ is the charge
of the radiator (throughout this paper the index $R$ refers to the radiators) and
\begin{equation}
u_{ei}( r) =\frac{1}{r}( 1-e^{-r/\delta}) ,
\label{eq:13}
\end{equation}%
which is regularized at small distances. The cutoff parameter $\delta$ may be qualitatively
thought of as a classical emulation of the electron thermal de Broglie length. For large
distances $r>\delta$ the potential becomes Coulomb, while for $r<\delta$ the Coulomb
singularity is removed and $u_{ei}(0)=1/\delta$. By this the short range effects based
on the uncertainty principle are included \cite{tal02,kel63,deu81}.

For a classical description of a plasma the electron degeneracy parameter $\Theta _{e}$,
i.e., the ratio of the thermal energy and the Fermi energy must fulfill
$\Theta _{e}=k_{B}T/E_{F}>1$. Since an ion is much heavier than an electron this condition
is usually fulfilled for ions. Therefore one can expect that the regularization given by
Eq.~\eqref{eq:13} is less important for ions than for electrons. Furthermore, scattering
of any two particles is classical for impact parameters that are large compared to the
de Broglie wavelengths $\lambdabar _{\alpha\beta}=( 2\pi\hbar^{2}/\mu_{\alpha\beta }k_{B}T)%
^{1/2}$, where $\mu_{\alpha\beta }$ is the reduced mass of the particles $\alpha $ and $\beta $.
Typical impact parameters are given by the Landau lengths,
$\lambda _{L\alpha \beta }=e_{S}^{2}\vert q_{\alpha }q_{\beta }\vert /k_{B}T$. Its ratio
to the de Broglie wavelengths is given by
\begin{equation}
\sigma _{\alpha \beta }=\frac{\Gamma _{ei} \left\vert q_{\alpha }q_{\beta }\right\vert }{Z}%
\frac{a}{\lambdabar _{\alpha \beta }} \ .
\label{eq:14}
\end{equation}%
Note that $\sigma _{ee}<\sigma _{ei}\ll \sigma _{ii}$. A classical description of
the scattering events in the TCP is valid if $\sigma _{ee}>1$.

A collective length scale is given by the Debye screening radius, for a TCP
$\lambda_{D}=a/(3\Gamma _{ei})^{1/2}$.
The plasma frequencies for electrons and ions $\omega _{p\alpha }=(4\pi q_{\alpha }^{2}%
n_{\alpha }e_{S}^{2}/m_{\alpha })^{1/2}$ with $\alpha =e,i$ set the collective time
scales $\omega _{pe}^{-1}$ and $\omega _{pi}^{-1}$ for electronic and ionic subspecies,
respectively. Due to their large mass ratio the electrons and the ions move on very different
timescales.
Moreover, for a nonrelativistic treatment the thermal energy of the
particles must be smaller than their rest energy. Since this is important only for
electrons we require that $k_{B}T\ll m_{e}c^{2}$. Also the validity of the dipole approximation
for plasma--radiator interaction used in Sec.~\ref{sec:3} requires that the characteristic
length scale of the plasma microfield must be larger than the effective atomic length scale
$a_{Z}=a_{B}/Z$, where $a_{B}$ is the Bohr radius. Since this length is $\simeq a$ the
dipole approximation is valid when $a\gtrsim a_{Z}$ which is usually fulfilled for heavy
radiators. In this paper we consider hydrogen--like ions as radiators in a completely
ionized TCP, i.e. we assume that $q_{R}=Z_{R}-1$, where $Z_{R}e$ is the charge of the
nucleus of the radiating ion.

Let us now briefly discuss the limitations of the MD model arising from the classical
description of the electrons, i.e., from the neglect of the quantum effects in the
short--range electron--ion interaction. As shown in Ref.~\cite{fis01} two constraints
on the parameter $\delta$ determining the regularized potential \eqref{eq:13} must be
considered. In the parameter regime when a significant fraction of the simulated electrons
is found in the quasibound states the simulated $Z$ and $Z_R$ are effectively reduced
and the MD simulations are not adequate. Thus, the parameter $\delta$ must be chosen
large enough to suppress the formation of the classical bound states of electrons. On
the other hand the condition $\delta\lesssim a$ must be fulfilled so not to affect the free
electron density at $r\sim a$. The probability to found an electron within a volume
$r\lesssim\delta $ from an ion is estimated by $W\simeq (4\pi\delta^{3} /3)n_{e}g_{ei}(0)$,
where the electron--ion radial distribution function $g_{ei}(r)$ can be approximated by
the nonlinear Debye--H\"{u}ckel expression $g_{ei}(r)\simeq \exp{[\beta_{e} Ze^{2}_{S}u_{ei}(r)]}$
(see, e.g., Refs.~\cite{ner05,tal02}) with $\beta_{e} =1/k_ {B}T$. Note that in the case
when $Z\gg 1$, the capture of an electron on the quasibound orbit reduce the effective
$Z$ and $\Gamma _{ei}$ for the next one, so a significant fraction of electrons can stay
free even for $\Gamma _{ei}a/\delta \gg 1$. The minimal value of $W(\delta )$ occurs at
$\delta =\Gamma_{ei}a/3$. It is thus clear that for $\Gamma_{ei} \gtrsim 0.5$ the capture
of the electrons onto classical orbits becomes important and the significant contribution
from the classically bound electrons cannot be avoided in the MD models with point--like
classical electrons.

\begin{figure*}[tbp]
\centering
\includegraphics[width=165mm]{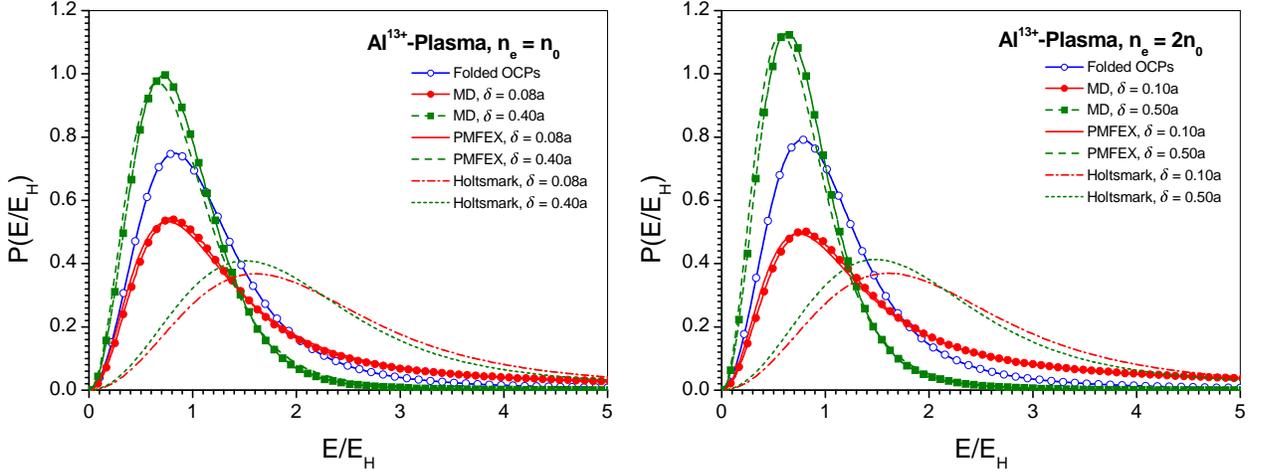}
\caption{Normalized electric microfield distributions for Al$^{13+}$ TCPs with
$k_{B}T=500$ eV, $n_{e}=n_{0}$, $\Gamma _{ee}=0.037$, $\Gamma _{ii}=2.65$ (left)
and $n_{e}=2n_{0}$, $\Gamma _{ee}=0.046$, $\Gamma _{ii}=3.34$ (right) as a function
of the electric field in units of $E_{H}$, Eq.~\eqref{eq:15}, for different values
of $\delta $. The lines with and without symbols represent the MFD from the MD
simulations and PMFEX model, respectively. The open circles are the MFD obtained
from the folding of an electronic and an ionic OCP, see Eq.~\eqref{eq:16}. The
Holtsmark distributions for a TCP (see Ref.~\cite{ner05}) are shown as dot--dashed
and dotted lines. (Online color:www.cpp-journal.org).}
\label{fig:1}
\end{figure*}

The electric microfield distribution (MFD) $P(E)$ plays a central
role for the line shape. Models for this distribution exist in the limits of an ideal
plasma \cite{hol19}, a weakly coupled plasma \cite{hoo68} and for very strongly coupled
plasmas \cite{may47}. For intermediate cases an effective independent--particle model
known as Adjustable Parameter Exponential (APEX) approximation has been developed in
Refs.~\cite{igl83,duf85,boe87} for an ionic OCP. It rests
essentially on the pair distribution function and has been tested by comparison with
MD and Monte--Carlo simulations. Recently in Refs.~\cite{ner05,ner06,ner08} we have
suggested the theoretical models named PMFEX (Potential of Mean Force Exponential
approximation) and PMFEX+ which turn out to be a very reliable approaches for calculating
the MFD of a TCP with attractive interaction. In order to cover the entire range from
small to large plasma parameters we use here classical MD simulations which have been
described in detail in Ref.~\cite{zwi99} (see also Refs.~\cite{ner05,ner06,ner08}). As
an example the normalized MFDs from PMFEX and MD are compared in Fig.~\ref{fig:1} where the electric microfields
are scaled in units of the Holtsmark field $E_{H}$ for a TCP (see \cite%
{ner05} for details)
\begin{equation}
E_{H}=\frac{C\mathcal{Z}e}{4\pi \varepsilon _{0}a^{2}},\qquad \mathcal{Z}=%
\left[ \frac{Z\left( 1+Z^{1/2}\right) }{Z+1}\right] ^{2/3}
\label{eq:15}
\end{equation}%
with an effective charge $\mathcal{Z}$ and $C=(8\pi /25)^{1/3}$. These distributions
were obtained from ensembles of fields taken at a charged reference point which is
chosen to be one of the plasma ions.

The MFDs for Al$^{13+}$ TCP with temperature $500$ eV and with coupling
parameters $\Gamma _{ee}=0.037$, $\Gamma _{ii}=2.65$
and $\Gamma _{ee}=0.046$, $\Gamma _{ii}=3.34$ are shown in left and right
panels of Fig.~\ref{fig:1}, respectively, for different values of $\delta$.
The density of plasma electrons is measured in units of $n_0
=5\times 10^{23}$~cm$^{-3}$. The dot--dashed and dotted curves are, for
comparison, the Holtsmark MFDs for a TCP with regularized Coulomb potential.
Here these MFDs depend on $Z$ and $\delta$ as discussed in \cite{ner05}. To
demonstrate the importance of the attractive interactions we also plotted
the MFDs $P_{0}(E)$ resulting from the corresponding electronic and ionic
OCPs with $\Gamma _{ee}$ and $\Gamma _{ii} $, respectively (open circles).
To that end the distribution $Q_{0}(\mathbf{E})=P_{0}(E)/(4\pi E^2)$ of the
total field $\mathbf{E}=\mathbf{E}_{1}+\mathbf{E}_{2}$ is calculated as
\begin{equation}
Q_{0}(\mathbf{E}) =\int d\mathbf{E}_{1}d\mathbf{E}_{2} \delta \left( \mathbf{%
E}-\mathbf{E}_{1}-\mathbf{E}_{2}\right) Q_{e} (\mathbf{E}_{1}) Q_{i} (%
\mathbf{E}_{2})
\label{eq:16}
\end{equation}%
from the MFD of the ionic OCP at a charged point $Q_{i}(\mathbf{E}_{2})$ and of
the electronic OCP at a neutral reference point $Q_{e}(\mathbf{E}_{1})$. The
distribution $Q_{0}(\mathbf{E})$ thus represents the MFD in a TCP assuming that
the ion--electron attractive interaction is absent. Here $Q_{e}(\mathbf{E}_{1})$
and $Q_{i}(\mathbf{E}_{2})$ are taken from MD simulations of an OCP. As the
thermal motion of the particles is suppressed with increasing coupling the
distributions $P(E)$ and the mean electric fields are shifted towards smaller
values as shown in Fig.~\ref{fig:1}. This figure also shows the importance of the attractive
interactions in plasmas. The behavior of the MFD with respect to the
variation of the parameter $\delta$ is particularly noteworthy. For fixed
coupling parameters the maximum of the MFD shifts only slightly to lower
field strengths $E$ with increasing $\delta$ (Fig.~\ref{fig:1}), while
the maximum itself increases with $\delta$. This is related to the largest
possible single--particle field $|E_e(0)| =e /(8\pi\varepsilon_{0}\delta^2 )$%
, which an electron can produce at the ion. Thus the nearest neighbor
electronic MFD vanishes for electric fields larger than $|E_e(0)|$, and
smaller $\delta$ will result in larger contributions to $P(E)$ at higher
fields $E$ with a corresponding reduction of $P(E)$ at small fields.
Therefore, the formation of the tails in the MFD and enhancement of the
electric microfield at small $\delta$ may have important influence on the
spectral line shapes of the radiating particles. Further examples for
charged and neutral radiators, together with a detailed discussion of the
limits of the PMFEX treatment at increasing coupling, are given in Refs.~%
\cite{ner05,ner06,ner08}.

\section{Wave equation for a radiator}
\label{sec:3}

In this section we describe the solution of the wave equation for a hydrogen--like
ion coupled to the time--dependent electric microfield. The microfield fluctuations
in the plasma are calculated by the MD simulations as discussed in Sec.~\ref{sec:2}.

We consider a hydrogen--like ion in a time--dependent electric microfield. The
Hamiltonian is the sum of $\hat{H}_{0}$ describing the unperturbed ion and a
dipole term $\hat{H}_{\mathrm{int}}=e\mathbf{r}\cdot \mathbf{E}(t)$ for the
interaction between the bound electron (distance $\mathbf{r}$ from nucleus) and
the microfield $\mathbf{E}(t)$,
\begin{equation}
\hat{H}=\hat{H}_{0}+\hat{H}_{\mathrm{int}} \ .
\label{eq:1}
\end{equation}%
The electron moves in the potential of a nucleus with charge $Z_{R}e$. In the present
application it turns out that it suffices to start from the non--relativistic
Schr\"{o}dinger equation
\begin{equation}
\hat{H}_{0}\left\vert \alpha \right\rangle =\left( \frac{\hat{\mathbf{p}}^{2}%
}{2m_{e}}-\frac{Z_{R}e_{S}^{2}}{r}\right) \left\vert \alpha \right\rangle =\hbar
\omega _{\alpha }\left\vert \alpha \right\rangle
\label{eq:2}
\end{equation}%
for the time--independent electronic state $\vert \alpha \rangle $ with energy
$E_{\alpha }=\hbar \omega _{\alpha }$. Here $\hat{\mathbf{p}}$ is the momentum
operator and $\alpha $ is a multiindex including radial, angular momentum and
spin quantum numbers. The present calculations are done in the configuration
space corresponding to the solutions of Eq.~\eqref{eq:2}. In order to discretize
the continuum a boundary condition $\langle \mathbf{r}|\alpha \rangle =0$ is
imposed at a radius $r=R_{0}$, which is chosen sufficiently large in order to
avoid an influence on the final results. The radial wave functions with this
boundary condition are still confluent hypergeometric functions, but the radial
quantum numbers of bound states are not integers any more \cite{mar03}.
In order to obtain a finite basis the (former) continuum states are cut off at
sufficiently large quantum numbers. Alternatively the continuum could be handled
by forming wave packets with a width that must be adjusted appropriately \cite{mul94}.
In Refs.~\cite{mar03,mar04} the time--dependent equation $\hat{H}\Psi (t)=%
i\hbar \dot{\Psi}(t)$ with the Hamiltonian \eqref{eq:1} has also been solved
on a grid for the electron wave function \cite{cal00,rei00}. This is more
advantageous for the description of the continuum and it is easier to implement
the interactions between the radiator and the plasma particles beyond the dipole
term in Eq.~\eqref{eq:1}. On the other hand spatially extended states require
very large simulation boxes. Quite generally in the present context the solution
on the grid is more expensive numerically than working in configuration space.
We adopted the latter for the subsequent calculations.

At high $Z_{R}$ relativistic corrections must be considered and also the spin should
be treated as a dynamical variable. It turns out that in the cases considered here it
suffices to include the first--order fine--structure shift \cite{fri90}
\begin{equation}
\Delta E_{nlj}=-\frac{Z_{R}^{2}\alpha _{S}^{2}\left\vert E_{n}\right\vert }{n%
}\left( \frac{1}{\gamma _{lj}}-\frac{3}{4n}\right) \ ,
\label{eq:3}
\end{equation}%
where $\gamma _{lj}=j+1/2$ and $\gamma _{lj}=1$ at $l\geqslant 1$ and $l=0$,
respectively. Here $n$ is the principal quantum number of the hydrogen--like
ion, $E_{n}=-Z_{R}^{2}E_{B}/n^{2}$ is the corresponding energy ($E_{B}=e_{S}^{2}/2a_{B}$
is the Bohr energy), $j=l\pm 1/2$ is the total angular momentum quantum number and
$\alpha _{S}\simeq 1/137$ is the Sommerfeld constant.

The dipole interaction $e\mathbf{r}\cdot \mathbf{E}(t)$ between the radiator and the
plasma is time--dependent and possibly strong. Going beyond the second order treatment
of Ref.~\cite{jun00} we use the interaction picture with the unperturbed basis states
given by Eq.~\eqref{eq:2}. The time--dependent Schr\"{o}dinger equation with the total
Hamiltonian (\ref{eq:1}) can be solved using Dirac's method. The perturbed electron wave
function is represented as a sum of wave functions of the unperturbed Hamiltonian with
time--dependent coefficients $c_{\alpha }(t)$
\begin{equation}
\Psi (t)=\sum_{\alpha }c_{\alpha }(t)e^{-i\omega _{\alpha }t}\left\vert
\alpha \right\rangle \ .
\label{eq:4}
\end{equation}%
A substitution of Eq.~\eqref{eq:4} into the time--dependent Schr\"{o}dinger equation
and orthogonality of the spatial wave functions, i.e., $\langle \alpha |\beta \rangle %
=\delta _{\alpha \beta }$, gives the set of coupled ordinary and linear differential
equations
\begin{equation}
\dot{c}_{\alpha }(t)=-\frac{ie}{\hbar }\mathbf{E}\left( t\right) \cdot
\sum_{\beta }e^{i\omega _{\alpha \beta }t}c_{\beta }(t)\left\langle \alpha
\left\vert \mathbf{r}\right\vert \beta \right\rangle
\label{eq:5}
\end{equation}%
which is solved iteratively. Here $\hbar \omega _{\alpha \beta }$ is the transition
energy between atomic states $\alpha $ and $\beta $, i.e., $\omega _{\alpha \beta } %
=\omega _{\alpha }-\omega _{\beta }$. Within the dipole approximation the transition
rate per unit time and energy interval $\mathcal{I}(\omega )$ for the emission of photons is
proportional to the power spectrum of the dipole operator \cite{jac98}. It is defined
as the square of the absolute value of the Fourier transform of the expectation value
of the dipole operator. Hence we introduce
\begin{equation}
\mathcal{I}\left( \omega \right) =\frac{2e_{S}^{2}\omega ^{3}}{3\pi c^{3}\hbar
^{2}\tau }\left\vert \int_{0}^{\tau }\mathbf{d}(t)e^{i\omega t}dt\right\vert ^{2}
\label{eq:6}
\end{equation}%
with $\tau \to \infty $ and the expectation value of the dipole moment
\begin{equation}
\mathbf{d}(t)=\sum_{\beta ,\alpha }e^{i\omega _{\beta \alpha }t}c_{\beta
}^{\ast }(t)c_{\alpha }(t)\left\langle \beta |\mathbf{r}|\alpha
\right\rangle .
\label{eq:7}
\end{equation}

Let us now consider a transition $\alpha \to g$ downwards to a state $\vert g\rangle $
which is nearly filled, i.e. $c_{\beta }(t)=\delta _{\beta g}$. Then the dipole moment
in Eq.~\eqref{eq:7} is calculated with respect to the state $\vert g\rangle $ and
\begin{equation}
\mathcal{I}\left( \omega \right) =\frac{2e_{S}^{2}\omega ^{3}}{3\pi c^{3}\hbar
^{2}\tau }\left\vert \sum_{\alpha }\left\langle g\left\vert \mathbf{r}%
\right\vert \alpha \right\rangle \int_{0}^{\tau }c_{\alpha }(t)e^{i\left(
\omega -\omega _{\alpha g}\right) t}dt\right\vert ^{2}.
\label{eq:8}
\end{equation}%

Our numerical model includes also the interaction $\hat{H}_{\mathrm{e\gamma }}$ of the
radiating electron with the emitted photons which, however, has been neglected in
Eq.~\eqref{eq:1}. In this case as shown in Refs.~\cite{mar03,mar04} the total radiated
power given in Eq.~\eqref{eq:8} is underestimated for the excited radiators where
$\vert c_{g}\vert ^{2}<1$. This can be compensated by dividing through the time--averaged
occupation probability of the lower state
\begin{equation}
\mathcal{I}\left( \omega \right) \to \frac{\mathcal{I}\left( \omega \right) }{%
\langle \left\vert c_{g}(t)\right\vert ^{2} \rangle _{t}} \ .
\label{eq:9}
\end{equation}%
The subsequent calculations will be done in this dipole power spectrum approximation
(DPSA). We have tested the validity of this approximation in the wide range of plasma
and radiator parameters by comparing explicitly $\mathcal{I}(\omega )$ with the spectrum
obtained from the total Hamiltonian $\hat{H}^{\prime }=\hat{H}_{0}+\hat{H}_{\mathrm{int}}%
+\hat{H}_{\mathrm{e\gamma }}$. We have found that in the parameter regime considered in
Sec.~\ref{sec:4} the DPSA is justified as the emission of radiation through the interaction
$\hat{H}_{\mathrm{e\gamma }}$ changes the occupation probabilities of the radiator's states
on a much slower scale than the fluctuating electric microfields. However, our preliminary
results show that the electron--photon interaction $\hat{H}_{\mathrm{e\gamma }}$ may have
an important contribution to the wings of the line especially in the case of light emitters.
We intend to take up further studies on this issue in a separate paper.

At this stage we have neglected the feedback of the radiator's excitation to the plasma.
In this respect the plasma particles move as if they had an infinite mass. As they have
a finite velocity it appears as if the radiating electron is embedded in a plasma of
infinite temperature. Accordingly the time evolution of the total system will lead to an
equal population of all electronic states. As the time--dependent feedback could be
implemented only at a very great expense in the MD simulations we enforce a canonical
equilibrium state of the plasma and the radiating electron by modifying the interaction
in Eq.~\eqref{eq:5} according to
\begin{equation}
e\mathbf{r}\cdot \mathbf{E}(t)\rightarrow e^{-\beta_{e} \hat{H}_{0}/2}e\mathbf{r}%
\cdot \mathbf{E}(t)e^{\beta_{e} \hat{H}_{0}/2} .
\label{eq:10}
\end{equation}%
The time--dependent equation \eqref{eq:5} describing the coupling of the
microfield to the radiator is then solved for an ensemble of typically
thirty independent microfields which yields the mean emission as well as the
statistical error.

\section{Results}
\label{sec:4}

Using the theoretical background introduced so far we present in this Section
calculations of the shape of the Ly$_{\alpha }$--line of Al$^{12+}$ radiating
ion embedded in a Al$^{13+}$--TCP in a wide range of plasma parameters. As we
have mentioned in Sec.~\ref{sec:3} for heavy ions the relativistic corrections,
i.e. the fine structure of the levels must be accounted for. Using the fine
structure shift in Eq.~\eqref{eq:3} the unperturbed Ly$_{\alpha }$ transition
energy becomes
\begin{equation}
\hbar \omega _{\mathrm{Ly}_{\alpha }}=\frac{3}{4}Z_{R}^{2}E_{B}\left(
1+C_{j}Z_{R}^{2}\alpha _{S}^{2}\right)
\label{eq:Ly}
\end{equation}%
with $C_{1/2}=\frac{11}{48}$, $C_{3/2}=\frac{5}{16}$ and for Al$^{12+}$
ions $\hbar\omega_{\mathrm{Ly}_{\alpha }}\simeq 1728.1$ eV and $\hbar \omega _{\mathrm{Ly}%
_{\alpha }}\simeq 1729.4$ eV with $j=1/2$ and $j=3/2$, respectively.

We start from the line as it is broadened by the Al$^{13+}$ ions and
electrons in the plasma. Then we fold with the weighted fine structure shift
Eq.~\eqref{eq:3} which is $\hbar \Delta \omega \simeq 1.29$~eV according to
\begin{equation}
\mathcal{I}_{\mathrm{LS}}(\omega )  =\frac{1}{3}\mathcal{I}\left( \omega +\frac{2}{3}\Delta \omega
\right) +\frac{2}{3}\mathcal{I}\left( \omega -\frac{1}{3}\Delta \omega \right)
\label{eq:17}
\end{equation}%
and account for the Doppler effect \cite{gri97,sal98} which broadens a line with
the unperturbed frequency $\omega _{0}$ according to a Gaussian distribution
\begin{equation}
\mathcal{D}\left( \Delta \omega \right) =\frac{1}{\sqrt{2\pi }\sigma }\exp %
\left[ -\frac{1}{2}\left( \frac{\Delta \omega }{\sigma }\right) ^{2}\right] ,
\label{eq:18}
\end{equation}%
where $\Delta \omega =\omega -\omega _{0}$ and
\begin{equation}
\sigma ^{2}=\frac{\omega _{W}^{2}}{8\ln 2}=\omega _{0}^{2}\frac{k_{B}T}{%
M_{R}c^{2}}.
\label{eq:19}
\end{equation}%
Here $\hbar \omega _{W}$ is the full width at half maximum (FWHM) of the line and
$M_{R}$ is the radiating ion mass. Some values of FWHM $\hbar\omega _{W}$ for
aluminum are shown in Table~\ref{tab:1}. Finally, we fold to account for the
experimental resolution.

\begin{table}[tbp]
\caption{Doppler broadening (FWHM) of the Ly$_{\alpha }$--line in an
Al plasma.}
\label{tab:1}
\begin{center}
\begin{tabular}{ccccccc}
\hline\hline
$k_{B}T$ (eV) & \quad 10 & \quad 10$^{2}$ & \quad $5\times 10^{2}$ & \quad 10%
$^{3}$ & \quad 10$^{4}$ & \quad $10^{5}$ \\ \hline
FWHM (eV) & \quad 0.08 & \quad 0.26 & \quad 0.57 & \quad 0.81 & \quad 2.57 &
\quad 8.12 \\ \hline\hline
\end{tabular}%
\end{center}
\end{table}

\begin{figure}[tbp]
\centering
\includegraphics[width=80mm]{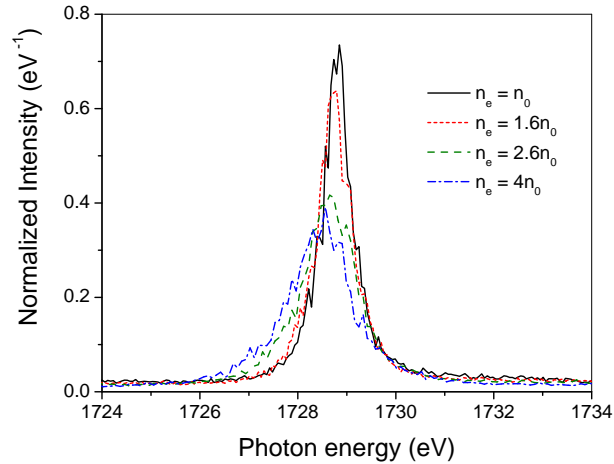}
\caption{Simulated Ly$_{\alpha }$--spectra of a Al$^{12+}$ radiating
ion embedded in a Al$^{13+}$-TCP of a temperature of 500~eV and solid state
densities $n_{e}=n_{0}$ (solid line), $n_{e}=1.6n_{0}$ (dotted line), $%
n_{e}=2.6n_{0}$ (dashed line) and $n_{e}=4n_{0}$ (dash-dotted line). The
spectra are normalized to the area under the curves. (Online color:www.cpp-journal.org).}
\label{fig:2}
\end{figure}

\begin{figure}[tbp]
\centering
\includegraphics[width=80mm]{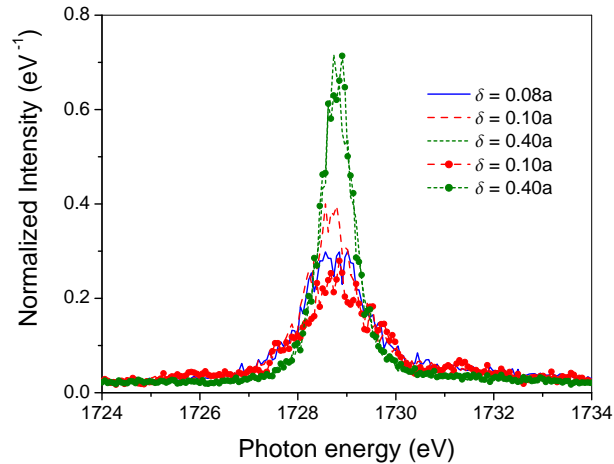}
\caption{Simulated Ly$_{\alpha }$--spectra of a Al$^{12+}$ radiating
ion embedded in a Al$^{13+}$-TCP of a density of $n_{e}=n_{0}$. The lines
without and with symbols correspond to the equilibrium ($T_{e}=T_{i}=500$
eV) and non-equilibrium ($T_{e}=500$ eV, $T_{i}=50$ eV) TCPs, respectively.
The regularization parameter is $\delta =0.08a$ (solid line), $\delta =0.1a$
(dashed lines) and $\delta =0.4a$ (dotted lines). The spectra are normalized
to the area under the curves. (Online color:www.cpp-journal.org).}
\label{fig:3}
\end{figure}

\begin{figure*}[tbp]
\centering
\includegraphics[width=165mm]{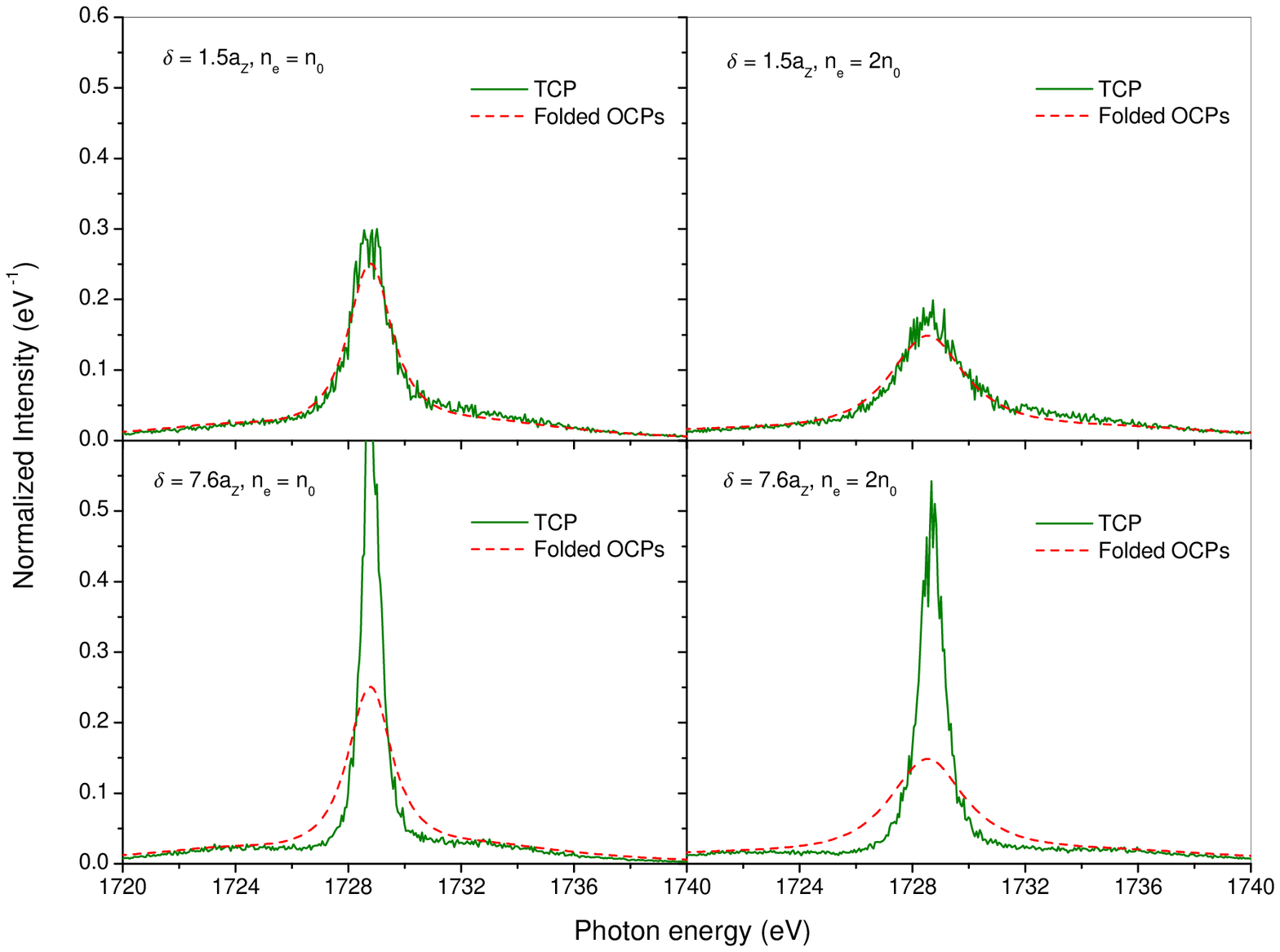}
\caption{Simulated Ly$_{\alpha }$--spectra of a Al$^{12+}$ radiating
ion embedded in a Al$^{13+}$--TCP (solid lines) of a temperature of 500~eV
and solid state densities $n_{e}=n_{0}$ (left panels) and $n_{e}=2n_{0}$
(right panels). The spectra are normalized to the area under the curves. The
parameter $\delta $ is $1.5a_{Z}$ and $7.6a_{Z}$ in the top and
bottom panels, respectively. The dashed lines are the spectra obtained from
the folding of an electronic and an ionic OCP. (Online color:www.cpp-journal.org).}
\label{fig:4}
\end{figure*}

\begin{figure}[tbp]
\centering
\includegraphics[width=80mm]{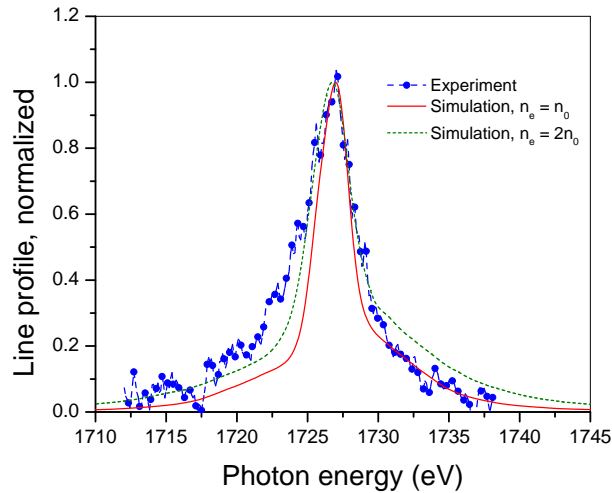}
\caption{Comparison of the experimental line \cite{and02} (filled
circles) with our simulation results (dashed and solid lines), i.e. with the
solid curves from the top panels of Fig.~\ref{fig:4} after taking
into account the Doppler broadening, the LS-coupling, the experimental
resolution and a redshift (see text). The experimental line is subtracted by
the underground. Here the curves are normalized to the peak intensity. (Online color:www.cpp-journal.org).}
\label{fig:5}
\end{figure}

We discuss now the simulated Ly$_{\alpha }$--line profiles at solid state densities
$n_{0}\leqslant n_{e}\leqslant 4n_{0}$ and at $k_{B}T=500$~eV. Some results for the
Ly$_{\alpha }$--line shape without Doppler broadening and LS coupling are shown in
Figs.~\ref{fig:2}-\ref{fig:5}. In all cases the ratio $a_{Z}/a\ll 1$ is small and the
use of the dipole approximation in Eq.~\eqref{eq:1} is fulfilled. The line broadening
and shift towards lower photon energies (redshift) is clearly visible in Fig.~\ref{fig:2},
which shows the line profile at fixed temperature 500~eV and different densities. Here
the regularization parameter $\delta =\lambdabar_{ei}$ is determined as the
thermal wavelength. With increasing density the influence of the plasma effects on the
line shape becomes more pronounced. At very large densities from $n_{e}=2.6n_{0}$
(dashed line) and up to the value $n_{e}=4n_{0}$ (dash--dotted line) there is hardly
any broadening and the line is only redshifted towards lower photon energies by the
plasma effects. In this high density regime the fluctuating electric fields become
sufficiently strong to cause asymmetric shapes due to nonlinear coupling.

To gain more insight we now fix the plasma temperature (500~eV) and the density
($n_{e}=n_{0}$) and show in Fig.~\ref{fig:3} the line profile for different
regularization parameters $\delta $ (the lines without symbols),
$\delta =0.08a$ (solid line), $\delta =0.1a$ (dashed line) and $\delta =0.4a$
(dotted line). Here the thermal wavelengths of the electrons are chosen
as the relevant lengths $\delta $ at which a smoothing of the ion--electron
interaction due to quantum diffraction becomes effective. For comparison we also
calculate the line shapes for non--isothermic plasma with different electronic
and ionic temperatures $T_{e}=500$~eV and $T_{i}=50$~eV, respectively (the lines
with symbols). As shown in Fig.~\ref{fig:3} the width of the lines decreases with
increasing parameter $\delta $, i.e. by "softening" of the ion--electron interaction.
Besides, keeping the electron temperature unchanged and decreasing the ionic
temperature leads to an additional broadening of the lines and this is visible for
more Coulomb--like interactions with small $\delta $.

In Fig.~\ref{fig:4} we put together the results obtained for two values of plasma
densities ($n_{0}$ and $2n_{0}$) and the parameter $\delta $. In the top and bottom
panels we take $\delta =1.5a_{Z}$ and $\delta =7.6a_{Z}$, respectively, where $a_{Z}$
is the effective Bohr radius of Al$^{12+}$. Note that for two plasma densities $n_{0}$
and $2n_{0}$ the chosen values of $\delta$ in units of $a_Z$ are equivalent to $0.08a$,
$0.1a$ and $0.4a$, $0.5a $, respectively, in units of the Wigner--Seitz radius $a$
of a TCP. We also demonstrate the influence of the electron--ion
attractive interaction on the spectral line shapes plotting the spectra $\mathcal{I}_{0}(\omega)$
resulting from a superposition of the electronic and ionic OCPs (dashed lines).
$\mathcal{I}_{0}(\omega)$ is calculated by folding the spectra $\mathcal{I}_{e}(\omega)$ and $\mathcal{I}_{i}(\omega)$
which are obtained from simulations of the radiative transitions of a radiator embedded
in an electronic OCP and of the ionic OCP, respectively. The microfields $\mathbf{E}(t)$
in an electronic and ionic OCPs are simulated at a neutral and charged reference points,
respectively. The spectrum $\mathcal{I}_{0}(\omega)$ thus represents the line shape in a TCP
assuming that the ion--electron attractive interaction is switched off. As shown in
Fig.~\ref{fig:4} the width of the line now turns out to be highly sensitive to both
the choice of $\delta $ and the density $n_{e}$, with a much stronger dependence on
$n_{e}$ for smaller $\delta $. The influence of the high--electric field tails in the
MFDs at small $\delta$ on the spectral lines is now clearly
shown in Fig.~\ref{fig:4}. Smaller $\delta$ results in higher electric fields which
broaden the spectral lines and reduce the peak intensity. More precisely we observed
that the line width behave approximately as $\hbar\Delta\omega \sim n_{e}/\delta$.

In the following we will compare our calculations with the results of experiments
performed in Garching \cite{and02} where an Al plasma is created by the irradiation
of the target with laser pulses of 150~fs duration at an intensity of a few $10^{17}$
W/cm$^2$. The systematic investigations carried out in Refs.~\cite{mar03,mar04} show
that the standard (quasistatic and impact) approximations become doubtful if the plasma
density reaches that of the solid state. In the last years experiments have approached
this regime, see, e.g., \cite{sae99,eid00,and02}. We note that the theoretical model
discussed so far assume a homogeneous equilibrium plasma. Obviously this is not the
state in which the laser leaves the target after the irradiating pulse. In particular
self--absorption due to plasma inhomogeneities leads to an additional line broadening
which is difficult to analyze. Fortunately there has been considerable experimental
progress to reduce the self--absorption \cite{and02}.

Earlier experiments on the Ly$_{\alpha }$--line in Al$^{12+}$ at solid state density
\cite{sae99,eid00} were subject to self--absorption in the cooler and less dense surface
regions of the target. This can be prevented by using thin (to reduce absorption) target
layers with sharp boundaries (to enhance homogeneity). For that purpose a 25~nm Al target
layer was embedded in solid carbon at depths ranging from $d=25$~nm to $d=400$~nm \cite{and02}.
With increasing depth the expansion of the Al layer is suppressed and the homogeneity
of the Al plasma is improved. In Fig.~\ref{fig:5} we compare our simulations with the
experimental results (filled circles) for $d=400$~nm and $k_{B}T=500$~eV from which the
underground has been subtracted. The fine structure and the Doppler broadening are taken
into account as described above. Then the simulated Ly$_{\alpha }$--lines are folded with
the experimental resolution (0.9~eV, FWHM) and compared with the experimental line assuming
densities $5\times 10^{23}$~cm$^{-3}$ (dashed line) and $10^{24}$~cm$^{-3}$ (thin solid line).
At these two densities the plasma parameters for the Al$^{13+}$--TCP are $\Gamma _{ii}=2.65$,
$\Gamma _{ee}=0.04 $ and $\Gamma _{ii}=3.34$, $\Gamma _{ee}=0.05$, respectively. All curves
in Fig.~\ref{fig:5} are normalized to the peak intensity. Finally, the position of the
simulated line must be redshifted by 2~eV. This is the dense plasma line shift (DPLS)
Ref.~\cite{gri97} due to the screening of the electron--nucleus interaction by hot background
electrons. Assuming a Debye--screened interaction instead of the $r^{-1}$--Coulomb potential
first--order perturbation theory yields a shift of the required magnitude. A comparison
of the two simulated curves in Fig.~\ref{fig:5} allows to conclude that the remaining
uncertainty in the determination of the density of the target is of order $10^{23}$~cm$^{-3}$.
Our results show that the quantum mechanics of close electron--ion collisions is important
over and above the plasma redshift. If the quantum diffraction parameter $\delta $ is fixed
at physically reasonable values near the effective Bohr radius $a_{Z}$, our calculations
favor a somewhat larger density than $n_{0}$ as proposed in Ref.~\cite{and02}. Clearly a
more quantum mechanical treatment of the electron component in the plasmas is desirable.

\section{Conclusions}
\label{sec:5}

In this paper we have presented a model for spectral lines that works without
some assumptions which underly the conventional impact and quasi--static
approximations. In particular we (i) consider two--component plasma (TCP) with
attractive interactions between electrons and ions, (ii) account for the strong
Coulomb correlations between plasma particles, (iii) account for radiator states
including the continuum, which are not directly involved in the transition, (iv)
allow for a non--perturbative treatment.

We have compared our model with recent experiments on Al targets and found good
agreement for the Ly$_{\alpha}$--transitions. The more exact treatments beyond the
standard approximations will become highly desirable in connection with experiments
at higher densities and temperatures at the planned (X)FEL facilities.

A critical discussion of our results suggests further improvements. (i) The dipole
approximation in Eq.~\eqref{eq:1} for the interaction of the microfield with the
radiator suffices for the present experiments \cite{and02}. In even denser plasmas
one must account for close collisions between the radiator and the plasma particles
with a quadrupole term in the expansion of the interaction and finally with an exact
treatment \cite{jun00}. (ii) Relativistic and spin effects beyond the simple fine
structure given by Eq.~\eqref{eq:3} can be taken into account by treating the radiator
with the Dirac equation \cite{mul94}. (iii) The major He--like satellite is well
separated from the Ly$_{\alpha }$--line in the experiment \cite{and02}. However,
there will be closer satellites due to spectator electrons in higher configurations,
which may affect the "red" shoulder of the line. For spectators in the continuum
this effect merges into the DPLS. The satellites impose a challenge as they offer
an additional tool to determine the temperature of the plasma, see, e.g. Ref.~\cite{sal98}.
For that purpose one has to solve the multi--electron wave equation, for example in
the relativistic case the Dirac equation \cite{dya89}. (iv) At high densities of the
laser--produced plasmas the self--absorption is likely to occur. In general, the
consideration of radiative transfer in dense plasmas is involved. To estimate the
effect of the self--absorption in the calculation of the line profile one--dimensional
monolayer model can be used which has already been successfully applied in Ref.~\cite{lor08}
(see also references therein).

\begin{acknowledgments}
This work was supported by the Deutsche Forschungsgemeinschaft (DFG-TO
91/5-3), the Bundesministerium f\"ur Bildung und Forschung (BMBF, 06ER128)
and by the Gesellschaft f\"ur Schwerionenforschung (GSI, ER/TOE).
\end{acknowledgments}

\end{document}